\documentclass[sigconf,authorversion,nonacm]{acmart}
\usepackage{tabularx}

\begin{document}
\title{Reducing the Barriers to Entry for Foundation Model Training}

\author{Paolo Faraboschi}
    \affiliation{\institution{Hewlett Packard Labs} \country{USA}}
\author{Ellis Giles}
    \affiliation{\institution{Hewlett Packard Enterprise} \country{USA}}
\author{Justin Hotard}
    \affiliation{\institution{Hewlett Packard Enterprise} \country{USA}}
\author{Konstanty Owczarek} 
    \affiliation{\institution{Hewlett Packard Enterprise} \country{USA}}
\author{Andrew Wheeler}
    \affiliation{\institution{Hewlett Packard Labs} \country{USA}}
    
\renewcommand{\shortauthors}{Faraboschi, Giles, et al.}

\begin{abstract}
The world has recently witnessed an unprecedented acceleration in demands for Machine Learning and Artificial Intelligence applications. This spike in demand has imposed tremendous strain on the underlying technology stack in supply chain, GPU-accelerated hardware, software, datacenter power density, and energy consumption. If left on the current technological trajectory, future demands show insurmountable spending trends, further limiting market players, stifling innovation, and widening the technology gap.

To address these challenges, we propose a fundamental change in the AI training infrastructure throughout the technology ecosystem. The changes require advancements in supercomputing and novel AI training approaches, from high-end software to low-level hardware, microprocessor, and chip design, while advancing the energy efficiency required by a sustainable infrastructure.

This paper presents the analytical framework that quantitatively highlights the challenges and points to the opportunities to reduce the barriers to entry for training large language models.
\end{abstract}


\maketitle

\section{Introduction}

\begin{figure}
\includegraphics[width=.8\columnwidth]{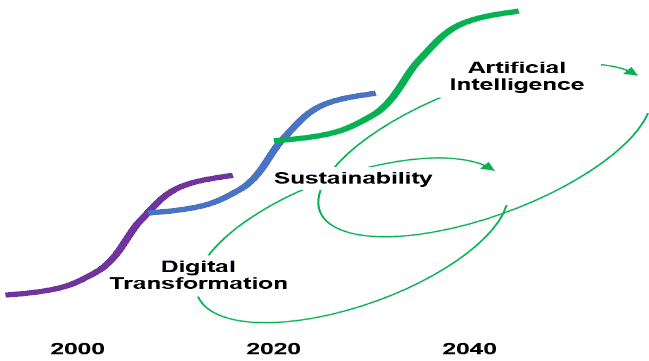}
\caption{Technology trends and AI}
\label{fig:1}
\end{figure}

Recent advances in intersecting technology areas from fast access to large data sets to accelerated computational power have driven compounding demand for increases in Machine Learning and Artificial Intelligence applications. This compounding effect has had implications in both immediate supply chain constraints (further exacerbated by pandemic effects), increases in cost, and future implications. A predicted S-Curve for AI adoption follows previous trends in Technology and Digital Transformation as depicted in Figure \ref{fig:1}.

Some sources \cite{Guardian23} compare the adoption of AI and impact on the worldwide population to that of the Industrial Revolution, while others compare Internet and Cloud Computing trends with historical digital adoption and transformation for business.
While the benefits of AI can have a substantial impact, it may come at a high cost.
For instance, future implications may include spending on AI applications and infrastructure that may exceed the current total IT spending.
Furthermore, if such a market remains concentrated, then only a few entities would be able to participate in such an expensive space. In this event, innovation may be stifled, and the technology gap widened.

To address these challenges, a fundamental change in the underlying architecture used to support the AI application landscape must be adopted and implemented throughout the worldwide technology ecosystem. Leaders in supercomputing technologies are well positioned to drive this change because of their track record and expertise in developing and deploying technologies at scale. Examples include software applications, development tools and libraries, and hardware from purpose-built silicon to datacenter-scale power and cooling all the while advancing high performing, energy efficient, and sustainable solutions. 

\subsection{The Landscape of Foundation Models}

Foundation models, such as large language models (LLMs) have revolutionized the field of natural language processing, enabling impressive advancements in text generation, comprehension, and a variety of language-related tasks. However, their development and usage are not without challenges. Training LLMs requires massive computational resources, including high-performance accelerated hardware, optimized interconnects and large-scale distributed systems which can be cost-prohibitive for many organizations. Collecting and cleaning the massive datasets of text (and code) to train LLMs is a daunting task, as it requires careful curation to avoid biases and errors.

These compounding challenges are causing the concentration of power of LLM training of proprietary models in the hands of a few players.
This near-monopoly poses several risks, potentially leading to a ``winner-takes-all'' scenario that stifles innovation and competition.
Additionally, it can also exacerbate societal inequality, by creating a divide between those who have access to those technologies, and those who do not.
The lack of transparency on how proprietary models are trained and what algorithms are used to rank the outputs can cause the propagation of misinformation or the emergence of biased censorship. The few players controlling LLMs also have access to a massive amount of sensitive personal information, which causes data privacy and security concerns.

One approach to mitigate these risks is regulatory, and several governments and influential organizations are taking steps to develop regulations that promote competition and innovation, data privacy and security protections, and ethical guidelines.

The complementary approach we propose advocates investing in technology that dramatically lowers the cost of foundation model training to reduce the barriers to entry for more players. For example, the open-source community has made significant contributions to the field of LLM training.
While the quality of open-source models has progressed and is similar to the commercial ones, the lack of access to affordable computation resources makes it difficult to train and serve open-source models at scale.
This difficulty in scaling creates challenges to the commercialization of open-source products and services, which is necessary to create the virtuous cycle to break the deadlock.

\subsection{LLM Growth Trends}

In the last few years, we have witnessed an unprecedented growth in the size of LLMs.
This remarkable growth can be attributed to multiple factors including access to more data (necessary to understand the language structure), access to powerful hardware (necessary to train the large models), and improved training algorithms (necessary to increase efficiency). In the last 3 years alone we have seen models double in size (number of parameters) every 4  months \cite{Sevilla22}, from 175B GPT-3 in 2020, 540B PaLM in 2022, and (rumored) 1,800B GPT-4 in 2023.
The community is divided on the future trajectory of LLMs, with some experts believing it will unavoidably slow down, and others expecting the growth to continue for many years to come, as researchers explore new applications and attack new data types beyond text.

In general, larger models have the potential to capture more intricate patterns and nuances in language.
These models exhibit improved expressiveness and a better understanding of context, allowing for more accurate and coherent responses. In tasks requiring a deep understanding of context or complex reasoning, a larger model performs far better in applications like question-answering, summarization, or dialogue systems. They also generalize better across a wide range of tasks and can deal with rare or uncommon language patterns. For example, the GPT-4 Technical Report \cite{OpenAI23}
shows the difference between GPT-3 (175B parameters) and GPT-4 (1,800B parameters, about 10x larger) on a set of standardized tests, where the larger GPT-4 is in the top 10\% of passing candidates vs. the smaller GPT-3 in the bottom 10\%.

There is also empirical evidence that we might see models over 100T parameters sooner than anticipated.
This 100T mark is particularly important because it is believed to be the order of magnitude of synapses (10\textsuperscript{11}) in the human brain \cite{Azevedo09}. Sparse training, such as Mixture-of-Experts (MoE \cite{Shazer17}) and Switch Transformers \cite{Fedus22}, is an active area of research and engineering and promises to yield an order of magnitude computation reduction, while preserving model quality. A team of researchers in China has recently disclosed details of BaGuaLu \cite{Ma22}, a 174 trillion parameter model which was trained on the 37 million cores of the Sunway supercomputer, showing advances in scaling possibilities.

Finally, several efforts are underway to expand foundation models beyond text and apply the same LLM principles  to areas such as video \cite{Ho22}, scientific research \cite{TPC23}, genomics \cite{Costa23}, and weather \cite{Patak22}. These new domains will require foundation models to grow in diverse ways, handling far larger context and different modes.

\subsection{Scaling LLM Training}

One of the biggest challenges of scaling LLM training as model size grows is the quadratic dependence of the training compute costs vs the number of parameters. In general, a neural model can be characterized by four parameters: size of the model in parameters (P), size of the training dataset in tokens (T), cost of training (in floating point operations, or FLOP, C), and performance after training. For simplicity, we are going to start with dense models, and defer discussing sparse models (such as MoEs) in the following sections. For transformer models, the foundation of modern LLMs, the ``Chinchilla scaling laws'' \cite{Hoffmann22} state that the training cost of an LLM (autoregressive, one epoch, cosine learning rate) is:

{\centering \textit{C = C}\textsubscript{\textit{0}} \textit{P T} \par}

Empirically, C\textsubscript{0} is approximately 6 FLOP per parameter, and the Chinchilla paper authors also found out that model size and the number of training tokens should be scaled in equal proportions. They evaluated that approximately 20 tokens per parameter are necessary for the model to converge. Putting this together, we get: 

{\centering \textit{C = 120 P}\textsuperscript{\textit{2}} \par}

For calibration, training a model with 1T parameters requires O(10\textsuperscript{26}) FLOP, which in most cases can be simplified to lower precision 16-bit multiplications and 32-bit accumulations. Considering a next-generation GPU performance of 1,000 16-bit TFLOP/s (10\textsuperscript{15} FLOP/s), it would ideally take 10\textsuperscript{11} GPU-seconds, which is approximately 28 million GPU-hours, to train that model. Inefficiencies at the algorithm, framework, and system levels get in the way of perfect scaling, and lead to an even far higher cost of training.

\begin{figure}
\includegraphics[width=1\columnwidth]{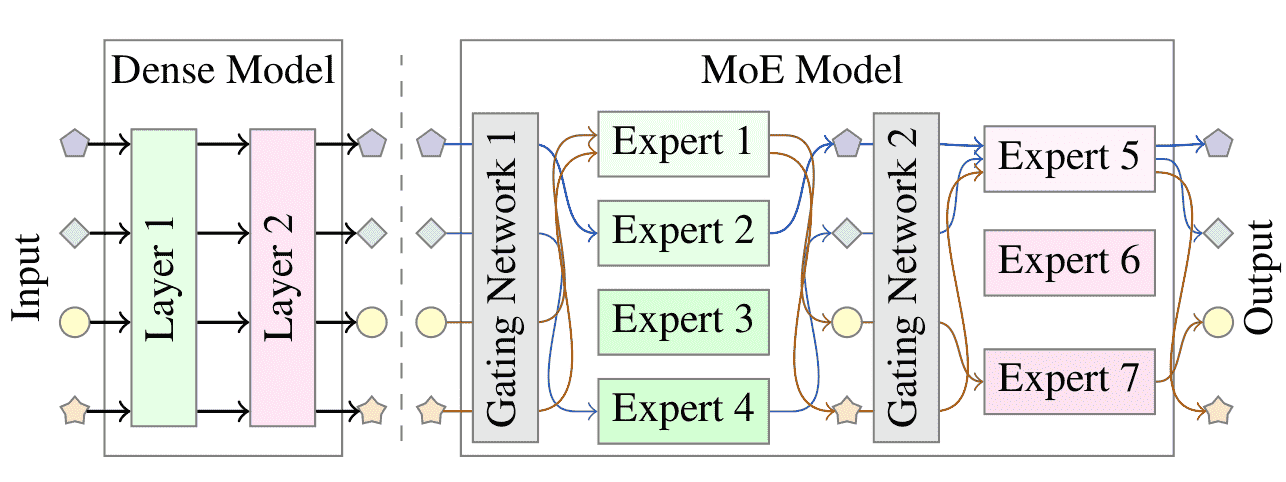}
\caption{Training of dense and sparse (MoE) models}
\label{fig:3}
\end{figure}

Recently, researchers have developed a variety of sparsely activated models, which in general fall under the category of ``Mixture-of-Experts,'' or MoE. Fig. \ref{fig:3} \cite{Zhai23} shows a conceptual diagram of the difference between dense and sparse models.

When training sparsely activated models, each training token only exercises a partition of the model, so the  cost is reduced. However, additional effort is required to traverse the routing networks that decide the specific model for each training token. The cost of training an MoE model with K experts can be approximated as

{ \centering \textit{C}\textsubscript{\textit{MoE}} \textit{= 120 K (P / K)}\textsuperscript{\textit{2}} \textit{= 120 P}\textsuperscript{\textit{2}} \textit{/ K} \par}

which corresponds to a cost reduction of a factor K.
Today, a typical range of K is 4 to 10 experts.  While research is exploring more aggressive approaches, it is difficult to imagine large K values due to gating and balancing.
The expert gating function, which determines the expert consulted, becomes more complex as K grows and becomes a sequential bottleneck.
Additionally, as K grows, the risk of experts becoming unbalanced becomes an issue, and the reduction in effective parallelism can negate the benefits of the lower training complexity.

\subsection{Proposed Approach}

We believe that the winning strategy to reduce the barriers to entry for LLM is to address \textit{model} \textit{producers,} because the prohibitive costs of model training drive most of the downstream ecosystem. Our analytical framework, discussed in Section II, confirms the exponential cost growth to train the largest AI models and the massive scale of the computational resources needed. These costs are outpacing the industry’s ability to deliver capacity through silicon and infrastructure innovation. 

The industry’s need for lower training costs, plus the momentum we are seeing in the market, points to several opportunities including:
\begin{itemize}
\item Adapting supercomputing technologies to LLM training, a 2x improvement to the cost of large-scale training is within reach soon. 
\item Empowering the \textit{open ecosystem} of open-source models enables everyone to build more efficient large-scale training systems and weakens the monopoly of proprietary models.
\item Delivering \textit{training-as-a-Service} at much lower cost removes further barriers to manage the complex compute, storage and networking infrastructure needed for training LLMs.
\end{itemize}
\subsection{Outline}

This paper analyzes  the LLM training challenges and outlines the opportunities that can mitigate the unsustainable cost explosion.

Section \ref{sec:af} presents our analytical framework which demonstrates the key challenges and importance and applicability of supercomputing technology as it relates to AI model complexity trained on a state-of-the-art system. As model sizes increase, the current strategy is focused on ``splitting'' them into sparsely activated partitions of mixture of experts (MoE). We believe that MoEs will still not close the gap between the cost to train AI models and the available global compute capacity available to train them.

Section \ref{sec:evo} discusses an ``evolutionary'' technical roadmap to achieve significant
cost reduction within the next 10 years. In the near term, a 2x supercomputing system technology can achieve reduction in the cost to train. In the medium and long term, we propose further optimizing an integrated solution where hardware and software are co-designed to reduce costs. 

Section \ref{sec:revo} presents a ``revolutionary'' approach that can further reduce the cost by moving away from deep learning and transformer models to an energy-based and physics-based approach that fundamentally changes the underlying paradigm.

We present our conclusions and next steps in Section \ref{sec:concl}.

\section{Analytical Framework \label{sec:af}}

\begin{figure}
\includegraphics[width=1\columnwidth]{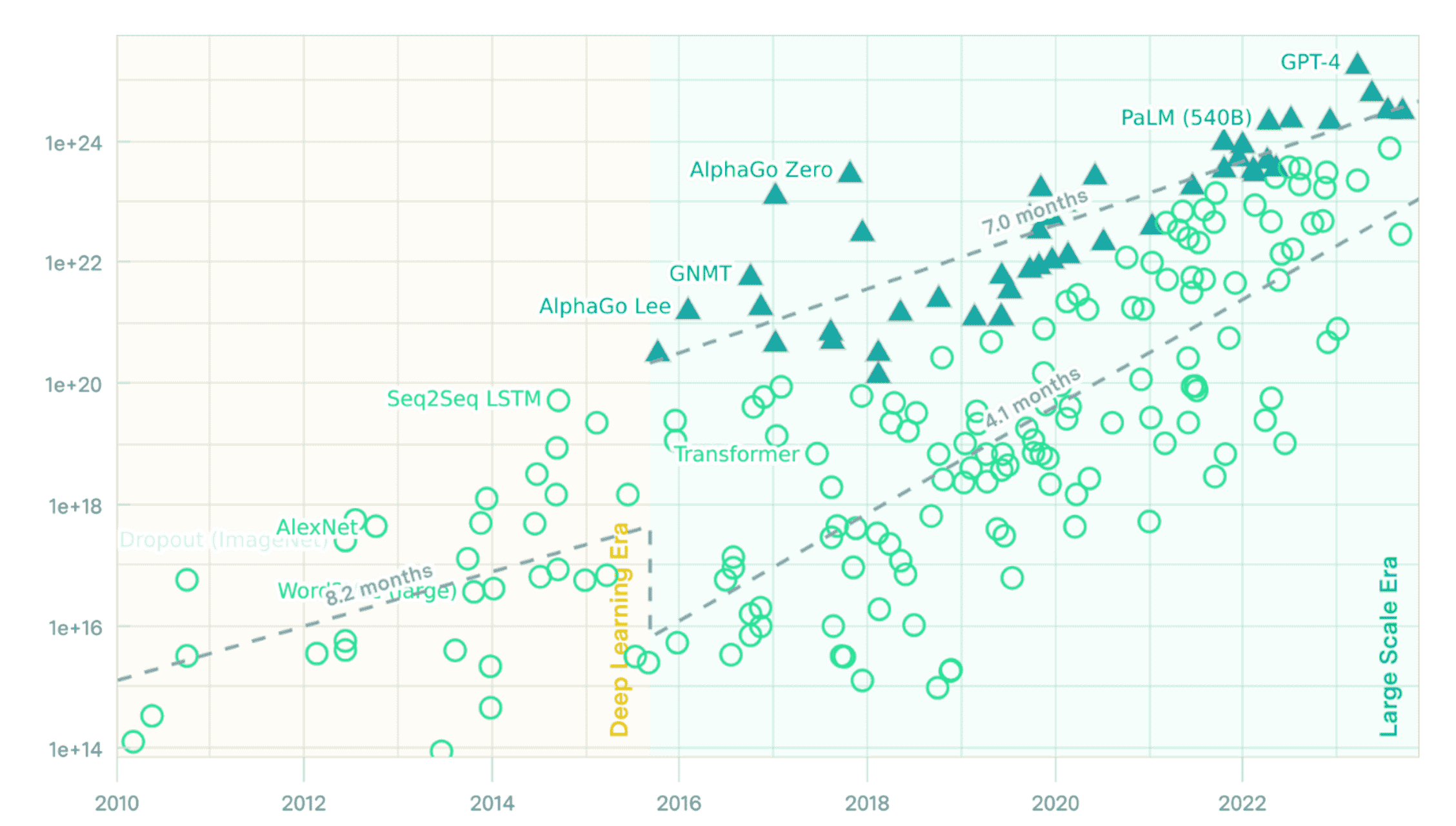}
\caption{Computational training cost trends in the deep learning and large scale era \cite{Epoch23}}
\label{fig:4}
\end{figure}

Researchers at Epoch AI (\url{https://epochai.org}) maintain a rigorous open database of ML systems and their computational complexity \cite{Sevilla22}. For example, Fig. \ref{fig:4} shows the computational cost of  notable LLMs in the last five years, pointing to a doubling time of approximately 4 months (or 7 months for large-scale models only) \cite{Epoch23}.

\begin{figure}
\includegraphics[width=1\columnwidth]{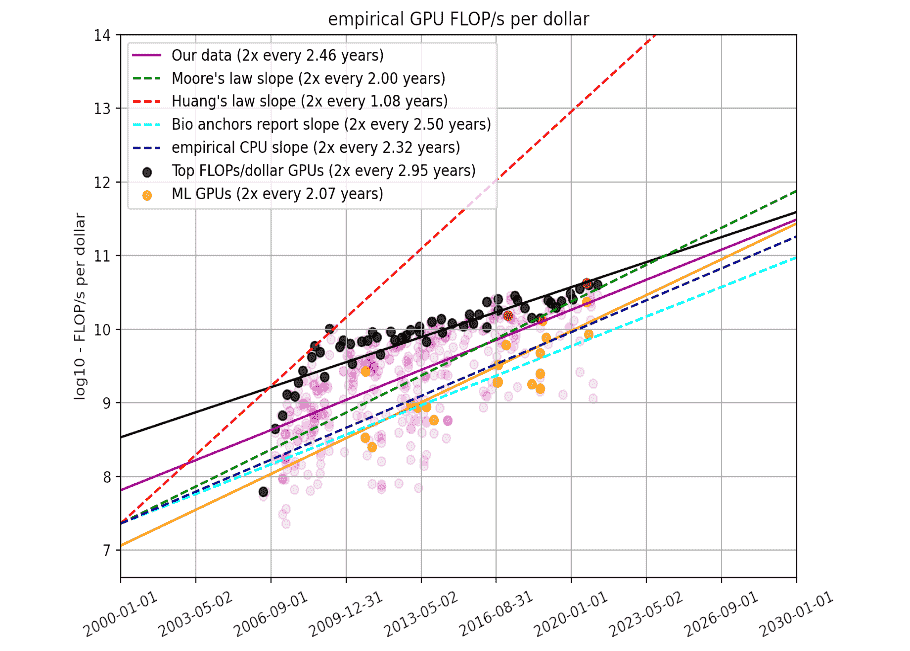}
\caption{GPU performance/\$ trends, last two decades \cite{Hobbhahn22}}
\label{fig:5}
\end{figure}

Unfortunately, the cost of LLM training has been growing at a far faster rate than the performance efficiency per unit price of GPUs. Fig. \ref{fig:5} \cite{Hobbhahn22} shows the GPU trends, expressed in FLOP/s/\$, of the last two decades, indicating a doubling time of approximately 2.46 years (29 months), which has been faster than the historical Moore’s Law slop, but still 4x-7x slower than the growth rate of the cost of LLM training shown in Fig. \ref{fig:4}.

\begin{figure*}[th]
\includegraphics[width=1\textwidth]{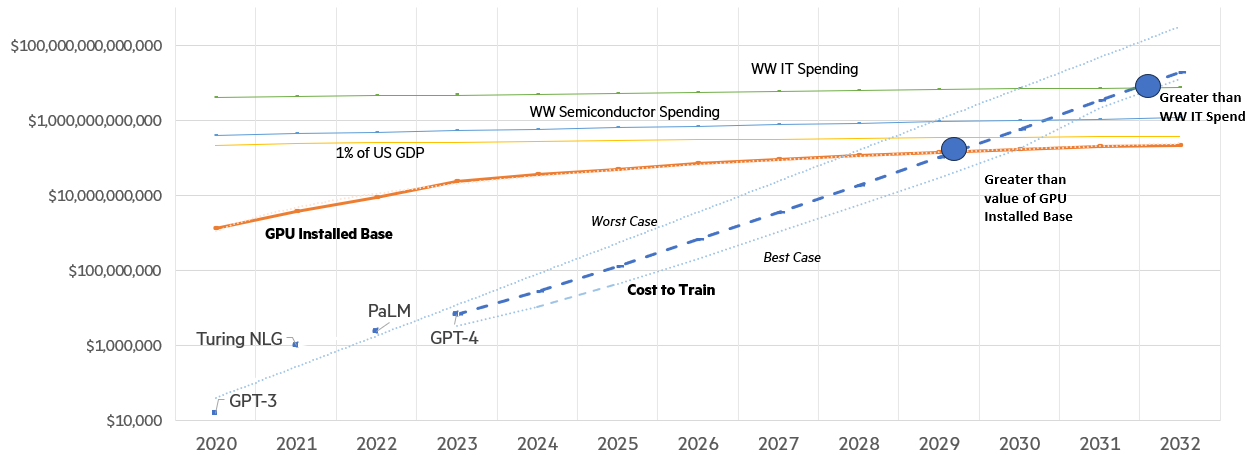}
\caption{Future projections for the cost of the final training run of a single LLM}
\label{fig:6}
\end{figure*}

Putting the two trends together implies that we are on a diverging path, where the cost of training is going to become increasingly unaffordable, even when granting GPUs an historical performance improvement which will slow down due to slower advances of the underlying silicon manufacturing process. This combination of trends is the cornerstone of our hypothesis that we are gradually but inexorably marching towards a ``winner-takes-all'' scenario, where fewer and fewer players will be able to afford training state-of-the-art models.

\subsection{Future Projections of LLM training cost}

To project future LLM training costs we have built a trend model that considers several factors, such as algorithmic improvements reducing the token/parameter ratio, and better sparsity models that increase the number of partitions without quality degradation.

Fig. \ref{fig:6} shows a projection of future LLM training costs as models continue to grow at a rate like what we have experienced in the last five years. We show three different projections: a ``best case'', a ``worst case'', and our ``best guess'' scenario. Table \ref{tab:1} shows the difference between the three scenarios

\begin{table}
\caption{ Best case, worst case, and best guess parameters}
\noindent\begin{tabularx}{\columnwidth}{|
p{\dimexpr 0.55\linewidth-2\tabcolsep-2\arrayrulewidth}|
p{\dimexpr 0.15\linewidth-2\tabcolsep-\arrayrulewidth}|
p{\dimexpr 0.15\linewidth-2\tabcolsep-\arrayrulewidth}|
p{\dimexpr 0.15\linewidth-2\tabcolsep-\arrayrulewidth}|} \hline 
 & Best case & Best guess & Worst case\\\hline 
Additional MoE experts/year & 8 & 4 & 0\\\hline 
FLOP/parameter (with tokens) & 20 & 40 & 120\\\hline 
\end{tabularx}
\label{tab:1}
\end{table}

As AI models significantly scale in size, supercomputing technology is a key enabler for efficient AI model training. The cost of training state-of-the-art models is growing rapidly and at an unsustainable pace, outpacing the growth in compute capacity. When GPT-3 (175B parameters) was trained in 2021, we estimate it took less than \$1M in GPU costs (corresponding to \$4.8M in cloud costs). By the time GPT-4 (estimated to be 1.8T parameters) was trained in 2023 the GPU cost had risen to \$6.7M (corresponding to \$32M in cloud costs). As we look to the future, and project out the demands of state-of-the-art models using past growth rates for model’s sizes (see  Figure 6), the cost to train a single model could rise to \$19B in 2028 in GPU costs alone (corresponding to \$93B in cloud costs).

At these growth rates, the cost to train a single LLM will exceed the annual value of the worldwide GPU installed base by 2029, and the annual global IT spending by 2032. This is despite forecasts for a rapidly growing GPU market and improvements in GPU performance per \$. This points to an industry situation where only a small group of players will be able to afford training the largest models. In this environment, these players would dictate norms for tuning and inference, and capture most of the economics. Even if the industry responds with increased compute supply, it will soon hit hard-to-break spending caps (e.g., 1\% of US GDP). Model training is becoming a winner-takes-all dynamic, which limits innovation.  Even when looking at best and worst case scenarios, the intersection points only move back or forward by no more than one year, which does not change the fundamental conclusion.

\begin{figure*}
\includegraphics[width=1\textwidth]{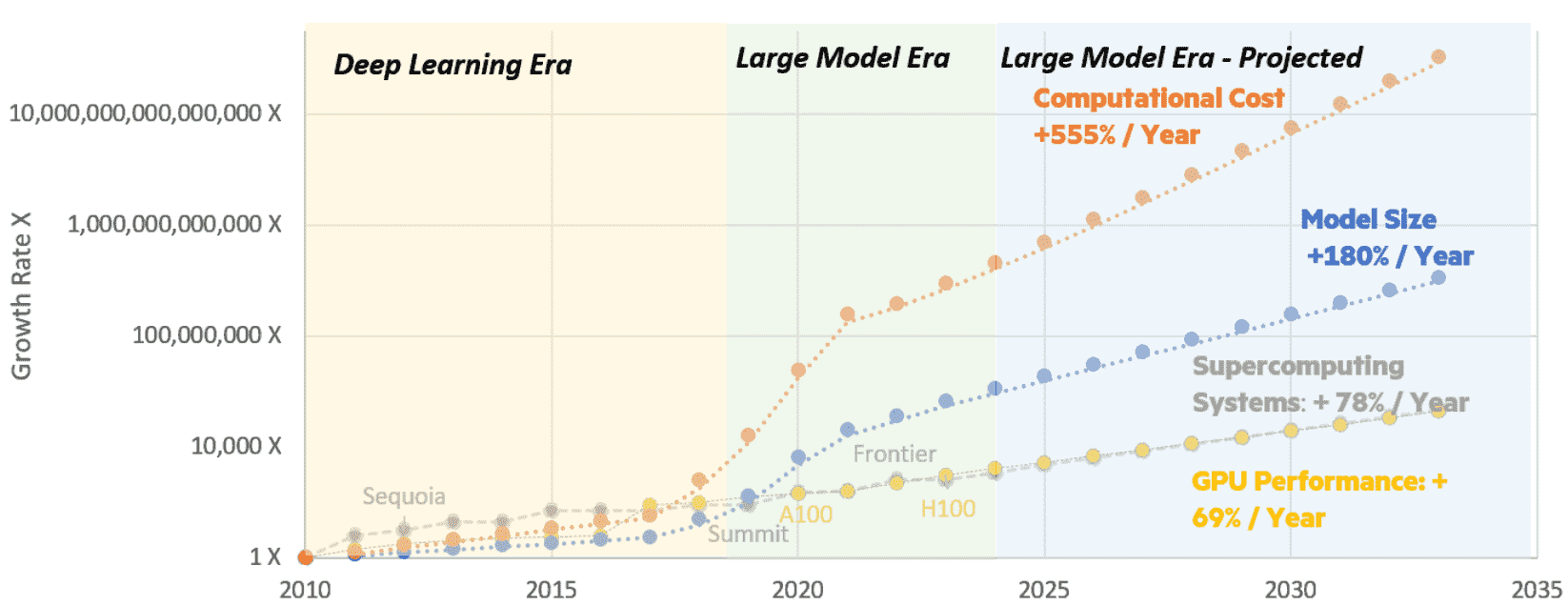}
\caption{Growth rate of GPU performance, model size and training computational costs (base in 2010)}
\label{fig:7}
\end{figure*}

As we previously discussed, this rapid growth in the cost to train a single state of the art model is driven by several factors. First, model sizes are growing exponentially. The number of parameters in these large models is forecasted to grow by 180\% per year with no end in sight, given the growth in underlying global data (e.g., the quadrillion sensor world) and AI creating its own data. The reason for this growth is also due to larger models showing a significant better quality of the generated results, as is evident even from the progression from GPT-2 to GPT-3 and GPT-4.

As described in previous sections, this situation is compounded by the quadratic relationship between model size and the computational cost required (in FLOP) to train these models. As a result (Fig. \ref{fig:7}), the number of FLOP required is growing at greater than 500\% per year. These computation requirements are growing by an order of magnitude greater than what the industry can deliver through improved GPU performance (\textasciitilde{}69\% per year) or the output of leading supercomputing systems (\textasciitilde{}78\% per year).

Before 2018, which marks the beginning of the ``\textit{Large Model Era}'', this was not the case; the industry was able to deliver increased performance to keep up with computational demands. However, the Large Model era changed this, and we have begun to see computational needs far outpace performance increases from next generation GPUs and supercomputers. 

\section{Evolutionary Roadmap \label{sec:evo}}

Identifying innovations to bend the cost curve on training the largest models is critical. Table \ref{tab:2} shows a list of technology areas and the 
description of the efficiency and improvement that is encompassed by each technology area.

\begin{table}[b]
\caption{Evolutionary Roadmap}
\noindent\begin{tabularx}{\columnwidth}{|
p{\dimexpr 0.24\linewidth-2\tabcolsep-2\arrayrulewidth}|
p{\dimexpr 0.76\linewidth-2\tabcolsep-\arrayrulewidth}|} \hline 
\centering\arraybackslash{}\textbf{Area} & \centering\arraybackslash{}\textbf{Description} \\\hline 
\centering\arraybackslash{}Distributed Systems & Increase system utilization with supercomputing technology \\\hline 
\centering\arraybackslash{}Data Reduction & Reduce tokens per parameter (regularization, lower dataset noise) \\\hline 
\centering\arraybackslash{}Model Partitioning & Increase sparsely activated partitions (mixture of experts) \\\hline 
\centering\arraybackslash{}Training Algorithms & Drive improvements that accelerate training convergence \\\hline 
\centering\arraybackslash{}Data Formats & Efficient numeric representations (sparsity, precision, logarithmic) \\\hline 
\centering\arraybackslash{}Efficient Hardware & Specialize beyond GPUs (remove fetch-decode overhead) \\\hline 
\centering\arraybackslash{}Competitive Accelerators & Align Accelerator margins to semiconductor averages \\\hline 
\end{tabularx}
\label{tab:2}
\end{table}

\subsection{Supercomputing Technology}

In distributed systems, supercomputing techniques will become increasingly important as system size increases to deliver cost savings. Supercomputing technology improves system-level resilience and effective utilization. That is done through a combination of techniques: more precisely identify the state to save, optimizing frequency and storage support for checkpointing, implementing a runtime that better tolerates certain failures. In our assumptions for ``optimized'' configuration, we are assuming a 2TB/s checkpointing performance, a 50\% reduction of checkpointing memory and a runtime that can tolerate 5 failed data parallel groups out of 100. 

\begin{figure}
\centering
\includegraphics[width=1\columnwidth]{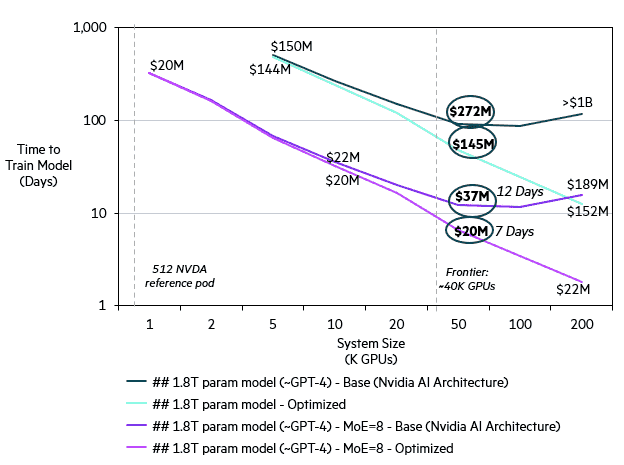}
\caption{Time to train one LLM vs system size (GPUs)}
\label{fig:8}
\end{figure}

The benefits of supercomputing technology improve as the system scales. With 50k GPUs we estimate that we can reduce the cost by \textasciitilde{}2x, less with smaller systems, more with larger systems. Without supercomputing technology, the performance of large systems no longer improves after a given size because the system spends most of its time attempting to recover from failures, rather than progressing the training job.

For example, a recent Google blog post \cite{Anantharaman23} presented details of the training run of a 32B parameter LLM over 50,944 TPU v5e accelerators. While the scaling is impressive, at 50k TPUs it is half the peak performance of the cluster, meaning that some scaling limitations start to appear in the cluster set-up.

In Fig. \ref{fig:8} we model two sets of curves to illustrate this point. The top two curves highlight the cost to train a single monolithic model, with and without supercomputing techniques. The lower two curves illustrate the benefits of organizing the model into eight sparsely activated partitions through the Mixture of Experts (MoE) approach, previously described.

Because of the quadratic relationship between model size and computational cost, breaking a model into smaller parts drives significant reduction in time and cost to train. However, this may degrade model quality and therefore may not be the right fit for every model. In both cases, applying supercomputing techniques drives \textasciitilde{}2x improvement. These techniques include software for resilience mitigation strategies optimized for large systems and long running jobs. This is also driven by the supercomputing experience identifying the optimal balance between hardware reliability and cost, and a fault-tolerant run-time system that uses optimized checkpoint-restart architectures.

The cost and system calculation are based on several assumptions, described in Table \ref{tab:3}.

\begin{table}
\caption{Parameters impacting training costs}
\noindent\begin{tabularx}{\columnwidth}{|
p{\dimexpr 0.3\linewidth-2\tabcolsep-2\arrayrulewidth}|
p{\dimexpr 0.13\linewidth-2\tabcolsep-\arrayrulewidth}|
p{\dimexpr 0.57\linewidth-2\tabcolsep-\arrayrulewidth}|} \hline 
GPU\_mem & 80 GB & GPU memory size,  for checkpoint time calculation\\\hline 
GPU\_mtbf\_base & 950k hours & GPU Mean Time Between Failures, from A100 datasheet \cite{A100}\\\hline 
CPU\_mtbf\_base & 1,500k hours & CPU Mean time between failures\\\hline 
GPUs\_per\_CPU & 4 & Number of GPUs for CPU for MTTI calculation\\\hline 
TFs\_per\_GPU & 150 TF32/s & Sustained 32b TFLOP/s (from NVidia A100 benchmarks)\\\hline 
Token\_scaling & 1.91 & Power scaling of parameter\newline (2.0 theoretical)\\\hline 
Token\_per\_param & 20 & Tokens per parameter\\\hline 
FLOP\_per\_token & 6 & Operations per token\\\hline 
Seq\_comp & 1\% & Sequential fraction of computation not parallelized (\%)\\\hline 
FT\_KN & {[}5,100{]} & Tolerated data parallel group failures (K-out-of-N)\\\hline 
TTR\_h = 2 & 2 hours & Time to recover a failed parallel group (hours)\\\hline 
FS\_base\_bw & 500 GB/s & Checkpointing filesystem base throughput\\\hline 
Cpgh & \$2.5 & Cost/GPU-hour (calibrated on dedicated A100 cloud instances)\\\hline 
\end{tabularx}
\label{tab:3}
\end{table}
\subsection{Data Reduction}

Data reduction techniques can help to lower the training costs of LLMs by reducing the size of the training dataset without sacrificing accuracy. Some of the promising approaches include Approximate Nearest Neighbor (ANN) search that remove redundant points in a high-dimensional space \cite{Tian23}, and dimensionality reduction to shrink dataset features through techniques such as Principal Component Analysis (PCA) \cite{Huertas23}. While these approaches have been previously studied in other contexts, their applicability to LLM training datasets is still in infancy and can yield significant reductions in cost.

\subsection{Model Partitioning}

The field of model \textit{sparsification} for LLMs is rapidly evolving. There is a wide range of promising techniques available, and new techniques are being developed all the time. In previous sections we described Mixture-of-Experts (MoE) and Switch Transformers that are in active development today, but other techniques, such as Routing Transformers \cite{Roy21} can reduce the overall complexity of an attention layer for sequence length \textit{n} and hidden dimension \textit{d} to O(n\textsuperscript{1.5}d) from O(n\textsuperscript{2}d). 

\subsection{Training Algorithms}

The majority of current training algorithms are based on a variation of Adam \cite{Kingma14} which has superseded the more traditional Stochastic Gradient Descent (SGD) used in the past. Adam has proven to converge faster than SGD, but it can be quite sensitive to hyper-parameter choices and can overshoot the optimal solution, leading to instability and slower convergence. Enhancements involve dynamically adjusting the learning rate of the model during training to improve the convergence speed and accuracy of the model and can potentially lead to a further reductions
in training costs.

\subsection{Data Formats}

LLMs are typically trained using lower-precision arithmetic, such as 16-bit or even 8-bit floating-point arithmetic. This can lead to significant speedups and memory savings, without sacrificing too much accuracy. There are a few reasons why LLM training works in lower precision arithmetic. First, LLMs are typically trained on large datasets of text and code. This means that there is a lot of redundancy in the data, which can be exploited by lower-precision arithmetic. For example, if a word appears multiple times in a training dataset, it is not necessary to store its full-precision representation each time. Instead, a lower-precision representation can be used, which can significantly reduce the memory footprint of the training dataset. GPUs have significantly taken advantage of these properties, including the use of mixed-precision training such as NVIDIA Transformer Engine \cite{TE23}.

New promising techniques for number formats are also appearing at the horizon. 

Shared block micro-exponents (SBME, \cite{Rouhani23}) is a floating point format that uses a shared exponent for a block of consecutive numbers. This can reduce the memory footprint of floating point numbers to an average as low as 3-4 bits, as the exponent only needs to be stored once for the entire block. SBME has been shown to be effective at reducing the memory footprint of LLMs by up to 50\%, without sacrificing performance. The format has recently been adopted by OCP as ``OCP Microscaling Format (MX)'' specification \cite{MX23}

Logarithmic number systems (LNS, \cite{Zhao22}) are a floating point format that represents numbers using their logarithms. This can simplify floating point operations because multiplications turn into additions in logarithmic space, and can lead to faster training times. LNS has been shown to reduce the training time of LLMs by up to 40\%, without sacrificing performance.

SBME and LNS are still under development, but they have the potential to further reduce the cost of LLM training.

\subsection{Efficient Hardware}

The last decade has witnessed a so-called ``Cambrian Explosion'' \cite{Farab18} of new AI/ML accelerators that promise to significantly improve the hardware efficiency. None of these approaches has so far emerged as the dominant alternative to GPU computing, but several promising directions are under development.

\begin{figure}
\includegraphics[width=1\columnwidth, height=2.1in]{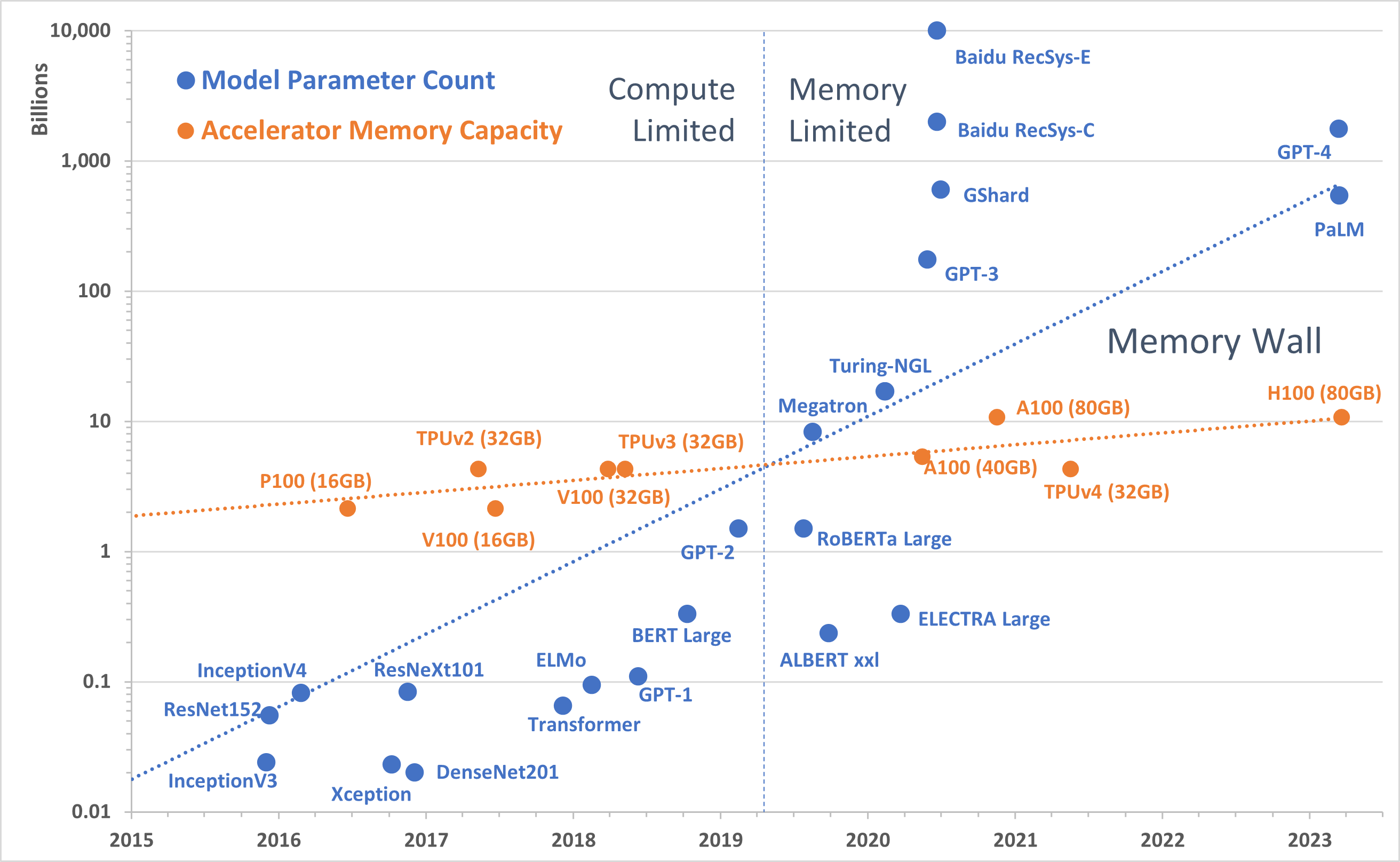}
\caption{The ``memory wall'' of LLMs \cite{gholami2020ai, celestial}}
\label{fig:9}
\vspace{-8pt}
\end{figure}

One of the key challenges of novel efficient hardware is addressing the ``memory wall'' caused by the mismatch of the growth rate of memory capacity and bandwidth vs. compute capacity \cite{gholami2020ai, celestial}.
As depicted in Figure \ref{fig:9}, the ``memory wall'' for training LLMs was reached in early 2019, when increasing model complexities caused their training to move from compute constrained to memory constrained.
To keep computing elements busy,  fast memories like HBM are needed to provision the required memory bandwidth. However, physical distance restrictions of HBM cause severe capacity issues, and today it is difficult to pack more than 100GB in a single package.

To overcome these limitations, photonics can decouple the memory bandwidth and capacity by disaggregating the fast memory tier without compromising the achievable bandwidth. Developments such as Celestial AI Photonics Fabric \cite{celestial} and Lightmatter Passage \cite{lightmatter} are exploring promising photonics-based approaches that can further reduce the training costs.

\subsection{Competitive Accelerators}

\begin{figure}
\includegraphics[width=.8\columnwidth, height=2.1in]{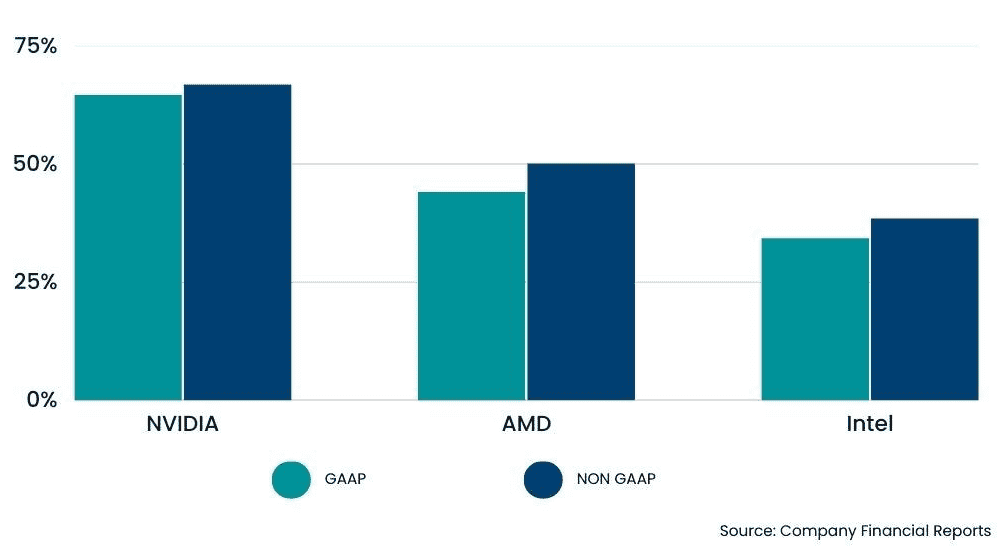}
\caption{Q1’24 gross margins for Nvidia, AMD and Intel}
\label{fig:10}
\vspace{-8pt}
\end{figure}

The rise of GPU-accelerated computing in the past decade has changed the data center chip market. The most important and expensive part of building a data center is no longer the CPU. Instead, large cloud and AI/ML companies are buying AI-accelerating GPUs which are today in such high demand that Nvidia’s datacenter revenue in August 2023 jumped 170\% \cite{Leswing23}, more than double the combined revenues of Intel and AMD. The positioning of Nvidia is also having significant effects on the GPU pricing and margins. Fig. \ref{fig:10} \cite{Jain23} outlines the gross profit margins of Nvidia (65\%), AMD (44\%) and Intel (35\%) in Q1’24 showing that market dominance and lack of competition has pushed Nvidia profit margins to exceptional levels. 
We believe that over time, the market will adjust itself, and increased competitive pressure of new accelerators will drive margins down to a historical Semiconductor range around 35\%-45\%, which can contribute to equivalent reductions in training costs.

\section{Revolutionary Roadmap \label{sec:revo}}

The dominant methods for training LLMs (described in the previous sections) use digital logic to perform the forward and backward passes of gradient descent. This approach is fundamentally limited by the speed at which data can be transferred between memory and compute units. Another key bottleneck of current AI models is that they are trained in the embedding space and not in their native formulation. In the embedding space, multiple auxiliary variables are created to represent the problem so that it can be mapped in nowadays accelerators, such as GPUs. A larger space means that the problem is larger (thus costly to map and accelerate in hardware) and more complex, since some extra difficulties can be generated by the auxiliary variables themselves. 

The AI/ML research community has been exploring radically different approaches for decades, which are now experiencing a resurging of interest due to the scaling issues of traditional approaches. Two promising fields are in-memory analog computing and energy-based models.
Combined, we believe they can show a potential to provide a further, drastic reduction in the cost of LLM training
However, several research breakthroughs in theory, algorithms and hardware implementations are still necessary to materialize these savings.

\subsection{Analog In-Memory Computing}

Analog In-Memory Computing (IMC) has the potential to overcome the memory bottleneck by performing computations directly on the memory, leading to significant speedups and energy savings. IMC has already shown applicability in other areas of AI, such as energy-efficient inference \cite{Ankit19} and explainable ML \cite{Pedretti21}.

The main idea is to implement both the fully connected layers and the attention computation (at the core of modern LLM architectures) in the analog domain but keep the readout digital, thus removing the costly conversion blocks required in a typical analog accelerator, while maintaining the massively parallel operations. We expect the same architecture can support both inference and training, but we envision two different memory technologies: non-volatile nanoscale memories (such as memristor ReRAM) are a good match for inference, while training requires analog CMOS memories with off the shelf components that can be scaled down to aggressive technology nodes. 

\subsection{Energy-Based Models}

An energy-based model (EBM) is a form of generative model derived from statistical physics which learns an underlying data distribution by analyzing a sample dataset. Once trained, an EBM can also produce other datasets matching the training data distribution \cite{EBM20}.

New, quantum-inspired, EBMs can automatically build a quantum field theoretic representation of the underlying probability distribution of a data source \cite{Huembeli22}. This statistical physics approach leads to interpretable emergent properties that can augment current generative models for more efficient and explainable performance. Characterizing different classes of expected and unexpected scaling laws and emergence properties in generative models could lead to completely new ways for training and inference.

EBMs have proven to be promising for solving constrained combinatorial optimization problems (k-SAT), where novel Hopfield Neural Networks (HNNs) with programmable high-order interactions between neurons map NP-hard problems into the most appropriate encoding, rather than force an embedding into a low-order (e.g., quadratic) space. A low-overhead, expandable, adjustable-topology circuit architecture allows to optimize the underlying platform for a particular problem set. Compact constraint enforcement circuits in the HNN loop result in an exponentially reduced solution search space. We can use mixed-signal memory circuits for quantum-inspired algorithms, and novel analog error-correction techniques to scale precision. 

Given that LLM training is essentially an optimization problem to find the best parameters to fit the model, we believe a similar concept can be applied to the training of LLMs in their native space. For example tokens in LLMs can be represented by their symbolic representation, preserving the relationship between them, without increasing the input space and effectively removing the dependence of training to model parameters. 

\section{Conclusions \label{sec:concl}}

In this paper, we have presented an analysis of the trends driving today’s large scale AI model training present a serious problem for the industry and society if left to grow unabated. In this work, we described the problem and provide an analytical framework which we use to judge the overall magnitude of the problem over the next decade given current trends. We then discussed several potential techniques, both revolutionary and evolutionary, which can be explored to mitigate these trends and foster open innovation. These techniques span the gamut of software and hardware approaches as well as their co-design, and each has an uncertain payoff over uncertain timelines. It is our opinion that organizations seeking to improve access to AI models will need to collectively invest in a subset of these techniques quickly while diligently scanning the horizon for new ones, otherwise the market dynamics will quickly move to a very concentrated set of winners.

\bibliographystyle{ACM-Reference-Format}
\bibliography{LLM.bib}

\end{document}